\documentclass[aps,pra,twocolumn,showpacs,superscriptaddress]{revtex4-1}

\usepackage{amsfonts,mathrsfs,amsmath,amsthm,amssymb}
\usepackage{bm,times}
\usepackage{graphicx,epsfig}
\usepackage[colorlinks=true,breaklinks=true,linkcolor=blue,citecolor=blue,urlcolor=blue]{hyperref}

\hypersetup{
  pdfauthor={Tao Zhou},
  pdfkeywords={},
  pdftitle={Performance of five-qubit code for arbitrary error channels},
  pdfsubject={Research Paper},
  pdfpagemode=UseNone
}

\def \tr{ {\rm{Tr}}}
\def \non {\nonumber}

\def \ket {\rangle}
\def \bra {\langle}

\newcommand{\be}{\begin{eqnarray}}
\newcommand{\ee}{\end{eqnarray}}
\newcommand{\dg}{\dagger}

\makeatletter
    
    \newcommand{\Rmnum}[1]{\expandafter\@slowromancap\romannumeral #1@}
\makeatother

\begin{document}
\title{Performance of five-qubit code for arbitrary error channels}
\author{Long Huang}
\affiliation{College of Physical Science and Technology, Sichuan University, Chengdu 610064, China}
\author{Xiaohua Wu}
\email{wxhscu@scu.edu.cn}
\affiliation{College of Physical Science and Technology, Sichuan University, Chengdu 610064, China}
\author{Tao Zhou}
\email{taozhou@swjtu.edu.cn}
\affiliation{School of Physical Science and Technology, Southwest Jiaotong University, Chengdu 610031, China}

\begin{abstract}
In quantum error correction, it is an important assumption that errors on different qubits are independent. In our previous work [Phys. Rev. A {\bf 92}, 052320 (2015)], the generality of the concatenated five-qubit code has been investgated when noises of the principal system and the auxiliary environment are assumed to be the same. In the error correction with concatenated code, tiny differences (in fidelity or independent errors) between initial quantum channels may introduce different effective channels in the next level, and therefore, it is necessary to study a meaningful question: Does five-qubit code still work efficiently when errors on different qubits are different? In the present work, it is discovered that even errors for different qubits are arbitrary, the five-qubit code still works efficiently. Since it is much easier and more accurate to measure the fidelity, one can construct quantum error correction with five-qubit code according to the initial channel fidelity, and it is not necessary to know the complete information of the initial channel. Moreover, when the initial channel fidelity is below $0.992$, the fidelity threshold for five-qubit code is the fidelity of the effective channel after error correction in bit-flip channels.
\end{abstract}

\pacs{ 03.67.Lx, 03.67.Pp}

\date{\today}

\maketitle

\section{Introduction}

In quantum computation and communication, quantum error correction (QEC) developed from classic schemes to  preserving coherent states from noise and other unexpected interactions. Shor~\cite{Shor} introduced a strategy to store a bit of quantum information in an entanglement state of nine qubits, and Steane~\cite{Steane} proposed a protocol that uses seven qubits. The five-qubit code was discovered by Bennett \textit{et al.}~\cite{Bennett} and independently by Laflamme \textit{et al.}~\cite{Laflamme}. Meanwhile, QEC conditions were proven independently by Bennett and co-authors~\cite{Bennett} and by Knill and Laflamme~\cite{KandL}. All the protocols with quantum error correction codes (QECCs) can be viewed as active error correction. Another way, the decoherence-free subspaces~\cite{Duan,Lidar,Zanardi} and noiseless subsystem~\cite{KandLV,Zanardi2,Kempe} are passive error-avoiding techniques.  Recently, it has been proven that both the active and passive QEC methods can be unified~\cite{Kribs,Poulin 05,Kribs2}.

The standard QEC procedure in Refs.~\cite{Steane,Bennett,Laflamme} is designed according to the principle of perfect correction for arbitrary single-qubit errors, where one postulates that single-qubit errors are the dominant terms in the noise process~\cite{Nielsen}. Recently, rather than correcting for arbitrary single-qubit errors, the error recovery scheme was adapted to model for the noise to maximize the fidelity of the operation~\cite{Reimpell,Fletcher 07,Yamamoto,Fletcher 08}. When the uncertainty of the noise channel is considered, robust channel-adapted QEC protocols have also been developed~\cite{Kosut PRL,Kosut,Ball¨®}. When the fidelity obtained from error correction is not high enough, the further increase in levels of concatenation is necessary. In the previous works~\cite{Poulin 06, Rahn, Kesting}, the concatenated code was discussed for the Pauli channel, where the depolarizing channel as the most important example is included. Before one can apply quantum error correction, we need to know the noise model to be corrected by measuring the Choi~matrix~\cite{Gilchrist,Emerson,Knill 08,Bendersky,Magesan 11,Magesan 12}. On the next level of error correction, the standard quantum process tomography (SQPT) \cite{Nielsen,D¡¯Ariano 03,D¡¯Ariano 04,Chuang,Poyatos,Leung} can be employed to determine the noise model of the effective channel. Finally, a nearly perfect quantum channel (with an error below $10^{-5}$) can be achieved via the concatenated quantum code.

In many models of quantum error correction, an important assumption--- the independence of errors on different qubits has been tested~\cite{Childs}. When errors on qubits are assumed to be the same, the generality of the concatenated five-qubit code has been investigated in Ref.~\cite{Huang}. When designing error correction with concatenated code, subtle differences (fidelity or the independent errors) between initial quantum channels may introduce different effective channels in the next level. Therefore, it is a meaningful and interesting question to study whether five-qubit code still work well when errors on different qubits are different.

In the present work, we utilize some common noise models to test five-qubit code, and meanwhile, the seven-qubit code is caculated with these common noise models as a contrast. The analytical results show that the five-qubit code is more appropriate than the seven-qubit code. Furthermore, the numerical calculation for five-qubit code has been performed when noise models are arbitrary, and the results show that the fidelity disparity between arbitrary channels and depolarizing channels is sufficiently small and the worst performance of five-qubit code is better than the performance of seven-qubit code in the chosen error channels. Moreover, the worst performance in five-qubit code occurs in bit-flip channels where the initial fidelity is below $0.992$. As shown in Ref.~\cite{Gilchrist,Emerson,Knill 08,Bendersky,Magesan 11,Magesan 12}, the average fidelity is much easier to measure than Choi~matrix, and thus, only the average fidelity about the initial channel is required in QEC, as long as five-qubit code works well in arbitrary error channels. If one takes the fidelity of effective channel after error correction in bit-flip error channels as the threshold, five-qubit code can be applied more widely.

The content of the present work is organized as follows. In Sec.~\ref{ii}, the entanglement fidelity and the standard quantum process tomography of quantum channel are briefly introduced. In Sec.~\ref{iii}, we will review the error correction protocol. In Sec.~\ref{iv}, the analytic results are performed when error models are bit flip, bit-phase flip, phase flip, amplitude damping and generalized amplitude damping in error correction with five-qubit code; In Sec.~\ref{v}, the analytic results in error correction with seven-qubit code are obtained when the errors in five-qubit code added another two depolarizing channels. In Sec.~\ref{vi}, the numerical calculation of error correction with five-qubit code is performed when error channels are generated arbitrarily. In Sec.~\ref{vii}, we end our paper with some remarks and discussion.

\section{Channel fidelity and standard quantum process tomography}
\label{ii}

Before one can design the protocol of error correction, it is necessary to know how well a quantun process $\varepsilon$ can store one qubit information, and usually, Schumacher's entanglement fidelity~\cite{Schumacher} can be used to describe it,
\begin{equation}
\label{b0def}
F=\langle S_{+}|\varepsilon\otimes I(|S_{+}\rangle\langle S_{+}|)|S_{+}\rangle.
\end{equation}
Here, $|S_{+}\rangle=1/\sqrt{2}(|00\ket+|11\ket)$ is a maximally entangled state. For a quantum channel $\varepsilon$, say
\be
\varepsilon(\rho)=\sum_{m}E_{m}\rho E^{\dagger}_{m},\ (m=0,1,2,3),\non
\ee
the entanglement fidelity $F$ can be expressed with the set of Kraus operators $\{E_{m},\ m=0,1,2,3\}$
\begin{equation}
\label{b1def}
F=\frac{1}{4}\sum_{m=0}^3(\tr E_m)^{2}.
\end{equation}
According to the result in Ref.~\cite{Horodecki}, entanglement fidelity $F$ has a beautiful relation with average fidelity $\bar{F}$,
\begin{equation}
\label{b2def}
\bar{F}=\frac{DF+1}{D+1}.
\end{equation}
In the present work, $D=2$, and in addition, the average fidelity $\bar{F}$ is defined as
\begin{equation}
\label{b3def}
\bar{F}=\int d\psi\langle\psi|\varepsilon(\psi)|\psi\rangle,
\end{equation}
where $d\psi$ is the Haar measure, say $\int d\psi=1$, and the integration is over the normalized state space. The fidelity averaged over the entire Hilbert space can be evaluated also by averaging over a state 2-design with a set of mutually unbiased basis~\cite{Bendersky}.

If one wants to obtain the complete information of the quantum process, the SQPT can be employed. The way of performing SQPT is not limited and in the present work, the protocol in Ref~\cite{X.-H. Wu} will be used. For a set of operators $E_{cd}=|c\ket\bra d|\ (c,d=0,1)$ as inputs for the principle system, the corresponding outputs are~\cite{Nielsen}
\be
\tilde{\varepsilon}(E_{cd})=\tr_\mathcal{A}[\mathcal{U}^{\dag}\circ\Lambda\circ\mathcal{U}(|a_0\rangle \langle a_0|\otimes|c\rangle\langle d|) ]\non
\ee
By introducing the coeficients
\begin{equation}
\label{b4def}
\tilde{\lambda}_{ab;cd}=\langle a|\tilde{\varepsilon}(E_{cd})|b\rangle,
\end{equation}
The Choi matrix of the effective channel $\tilde{\varepsilon}$ can be expanded as
\be
\chi(\tilde{\varepsilon})=\sum^{1}_{\emph{a},\emph{b},\emph{c},\emph{d}=0}\tilde{\chi}_{ab;cd}|ab\rangle\langle cd|,\non
\ee 
and the matrix elements are
\begin{equation}
\label{b5def}
\tilde{\chi}_{ab;cd}=\langle ab|\chi(\tilde{\varepsilon})|cd\rangle.
\end{equation}
It has been shown in \cite{X.-H. Wu}, $\tilde{\chi}_{ab;cd}$ can be obtained in a simple way,
\begin{equation}\label{b6def}
\tilde{\chi}_{ab;cd}=\tilde{\lambda}_{ac;bd}.
\end{equation}
Finally, one can come to a relationship between channel fidelity and elements of Choi matrix $\chi(\tilde{\varepsilon})$,
\begin{equation}\label{b7def}
F(\varepsilon)=\frac{1}{4}(\tilde{\chi}_{00;00}+\tilde{\chi}_{00;11}+\tilde{\chi}_{11;00}+\tilde{\chi}_{11;11})
\end{equation}

It is much easier and more accurate to measure the average fidelity than the Choi matrix of a quantum process~\cite{Emerson,Knill 08,Bendersky,Magesan 11,Magesan 12}, and therefore, it is enough to know the fidelity in the preparation of error correction.

\section{Unitary realization of quantum error correction}
\label{iii}

In this section, we will briefly introduce the realization of QEC. Following the idea in Refs.~\cite{Rahn,Kesting}, the exact performance of QEC can be quantified by the improvement of the channel fidelity, and this will make the calculation simplified~\cite{Huang}.

We start the QEC protocol with five-qubit code in Ref.~\cite{Bennett},
\be
|0_\mathcal{L}\rangle&=&\frac{1}{4}[|00000\rangle+|10010\rangle+|01001\rangle+|10100\rangle\non\\
&&+|01010\rangle-|11011\rangle-|00110\rangle-|11000\rangle\non\\
&&-|11101\rangle-|00011\rangle-|11110\rangle-|01111\rangle\non\\
&&-|10001\rangle-|01100\rangle-|10111\rangle+|00101\rangle],\non
\ee
and
\be
|1_\mathcal{L}\rangle&=&\frac{1}{4}[|11111\rangle+|01101\rangle+|10110\rangle+|01011\rangle\non\\
&&+|10101\rangle-|00100\rangle-|11001\rangle-|00111\rangle\non\\
&&-|00010\rangle-|11100\rangle-|00001\rangle-|10000\rangle\non\\
&&-|01110\rangle-|10011\rangle-|01000\rangle+|11010\rangle].\non
\ee

\begin{figure}[tbph]
\centering
\includegraphics[width=0.4 \textwidth]{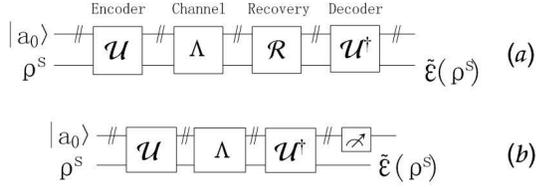}\\
\caption{(a) The way of getting the effective channel from the
standard QEC protocol includs encoding, noise evolution, recovery, and decoding.
 (b) Our protocol where the chosen unitary
transformation is sufficient to correct the errors of the principle
system. The errors of the ancilla system are left uncorrected. }\label{figure1}
\end{figure}

As depicted in Fig.~\ref{figure1}~(a), the standard way to get the effective noise channel contains the following steps: (i) A unitary transformation $U$ for encoding process $\mathcal{U}$; (ii) The noise evolution denoted by $\Lambda$; (iii) The recovery operation described by a process $\mathcal{R}$ such that ${\mathcal{R}(\rho^{\mathcal{SA}}})=\sum^{15}_{m=0}R_m\rho^{SA}R^{\dag}_{m}$, with $R_m$ the Kraus operators; (iv) The decoding process $\mathcal{U}^\dg$ realized by $U^{\dag}$.

When designing the QEC protocol in quantum computation, one can have
\be
\mathcal{R}\circ\mathcal{U}^\dg\equiv\mathcal{U}^\dg\circ\tilde{\mathcal{R}},\non
\ee
with a new process $\tilde{\mathcal{R}}=\mathcal{U}\circ\mathcal{R}\circ\mathcal{U}^\dg$. According to analysis in Ref.~\cite{Bennett}, the recovery process $\mathcal{\tilde{R}}$ is not necessary and can be moved away. If the protocol is applied in quantum information storage and transmission, the recovery of auxiliary qubits can be abandoned. In Fig.\ref{figure1} (b), a simplified protocol to obtain the effective channel is defined as
\begin{equation}
\label{adef}
\tilde{\varepsilon}(\rho ^{S})=\tr_\mathcal{A}[\mathcal{U}^{\dag}\circ\Lambda\circ\mathcal{U}(|a_0\rangle \langle a_0|\otimes\rho^\mathcal{S})].
\end{equation} 

In the protocol above, $\mathcal{U}$ and $\mathcal{U}^{\dagger}$ are the most important processes, where not only encoding and decoding, but also error correction is implemented. As designed in Ref.~\cite{Bennett} ,
\be
\label{a1def}
U|a_{m}\rangle\otimes|0\rangle&=&|m,+\rangle=E_{m}|0_\mathcal{L}\rangle,\non\\
U|a_{m}\rangle\otimes|1\rangle&=&|m,-\rangle=E_{m}|1_\mathcal{L}\rangle,
\ee
where $m=0,1,...,15$, $E_0$ is the identity operator $\hat{I}$, and for $m\neq0$, $E_{m}$ is one of the Pauli operators $\sigma^{i}_{j}(i=1,...5, j=x,y,z)$.

In the following, the Steane code~\cite{Steane} will be used in Sec.~\ref{v} as a contrast, and the logic states are
\be
|0_L\rangle&=&\frac{1}{2\sqrt{2}}[|0000000\rangle+|1010101\rangle+|0110011\rangle\non\\
&&+|1100110\rangle+|0001111\rangle+|1011010\rangle\non\\
&&+|0111100\rangle+|1101001\rangle],\non
\ee
and
\be
|1_L\rangle&=&\frac{1}{2\sqrt{2}}[|1111111\rangle+|0101010\rangle+|1001100\rangle\non\\
&&+|0011001\rangle+|1110000\rangle+|0100101\rangle\non\\
&&+|1000011\rangle+|0010110\rangle].\non
\ee

In the error correction with the steane code, encoding process $\mathcal{V}$ is a unitary transformation $V$ in a $2^7$-dimensional Hilbert space, and its inverse $V^\dg$ is the decoding process. The set of correctable errors $\{E_{m}\}^{63}_{m=0}$ consists of the identity operator $E_{0}=\hat{I}^{\otimes7}_{2}$, all the rank-one Pauli operators $\sigma^{j}_{i} (i=x,y,z, j=1,2,...,7)$ and a number of $42$ rank-two operators such as $\sigma^{1}_{x}\otimes\sigma^{2}_{y}, \sigma^{5}_{z}\otimes\sigma^{3}_{x},...$, etc.

With the definition
\be
|m,+\ket=E_m|0_L\ket,\ |m,-\ket=E_m|1_L\ket\non,
\ee
the set of normalized vectors $\{|m,\pm\ket\}_{m=0}^{63}$ constitutes a basis of the $2^7$-dimensional Hilbert space, and for the ancilla system, the basis is denoted by $\{|a_m\ket\}_{m=0}^{63}$. Then, from the requirements
\be
V|a_{m}\rangle\otimes|0\rangle&=&|m,+\rangle=E_{m}|0_{L}\rangle,\non\\
V|a_{m}\rangle\otimes|1\rangle&=&|m,-\rangle=E_{m}|1_{L}\rangle,\non
\ee
one can obtain the unitary transformation $V$ as
\be
V=\sum^{63}_{m=0}(|m,+\rangle\langle a_{m},0|+|m,-\rangle\langle a_{m},1|).\non
\ee
Finally, the effective channel for seven-qubit is
\begin{equation}
\label{a2def}
\tilde{\varepsilon}(\rho ^{S})=\tr_\mathcal{A}[\mathcal{V}^\dg\circ\Lambda\circ\mathcal{V}(|a_0\rangle \langle a_0|\otimes\rho ^{S}) ].
\end{equation}

In the selection of the correctable states, the orthogonality of correctable states in the process is required to perform effective correction capability for the seven-qubit code. The maxmum number for all the correctable errors is $m\ast g$, where $m$ is the number of the selected correctable states, also equal to the dimension of the Hilbert space divided by $2$, and $g$ is the number of generators for Steane code.

\section{Exact performance of five-qubit code for common error channels}
\label{iv}

In this section, to describe how the five-qubit code works against common error models, we take the depolarizing channel as the contrast and then compare the performance between common error channels and the depolarizing channels. The common errors considered here are bit flip error, bit-phase flip error, phase flip error, amplitude damping error and generalized amplitude damping error for five-qubit code. The initial channel fidelity $F_{0}$ is the probability that errors do not happen, and we set $F_{0}=p$ here. The Kraus operators can be expressed as,
\begin{equation}
E^{(1)}_\mathrm{BF}=\sqrt{p}\hat{I}_{2},\ E^{(2)}_\mathrm{BF}=\sqrt{1-p}\hat{\sigma}_{x}\non
\end{equation}
for bit flip channel,
\begin{equation}
E^{(1)}_\mathrm{BPF}=\sqrt{p}\hat{I}_{2},\ E^{(2)}_\mathrm{BPF}=\sqrt{1-p}\hat{\sigma}_{y}\non
\end{equation}
for bit-phase flip channel,
\begin{equation}
E^{(1)}_\mathrm{PF}=\sqrt{p}\hat{I}_{2},\ E^{(2)}_\mathrm{PF}=\sqrt{1-p}\hat{\sigma}_{z}.\non
\end{equation}
for phase flip channel,
\be
E^{(1)}_\mathrm{AD}=\left(\begin{array}{cc}
1 & 0\\
0 & P_1
\end{array}\right),
E^{(2)}_\mathrm{AD}=\left(\begin{array}{cc}
0 & P_2\\
0 & 0
\end{array}
\right).\non
\ee
for amplitude damping channel,
\be
E^{(1)}_\mathrm{GAD}&=&\sqrt{p}\left(\begin{array}{cc}
1 & 0\\
0 & P_1
\end{array}
\right),\
E^{(2)}_\mathrm{GAD}=\sqrt{p}\left(\begin{array}{cc}
0 & P_2\\
0 & 0
\end{array}\right),\non\\
E^{(3)}_\mathrm{GAD}&=&\sqrt{1-p}\left(\begin{array}{cc}
P_1 & 0\\
0 & 1\end{array}
\right),\
E^{(4)}_\mathrm{GAD}=\sqrt{1-p}\left(\begin{array}{cc}
0 & 0\\
P_2 & 0\end{array}
\right).\non
\ee
for generalized amplitude damping channel, with $P_1=|2\sqrt{p}-1|,P_2=\sqrt{4(\sqrt{p}-p)}$, and
\be
E^{(1)}_\mathrm{DEP}&=&\sqrt{p}\hat{I}_{2},\  E^{(2)}_\mathrm{DEP}=\sqrt{\frac{1-p}{3}}\hat{\sigma}_{x},\non\\
E^{(3)}_\mathrm{DEP}&=&\sqrt{\frac{1-p}{3}}\hat{\sigma}_{y},\ E^{(4)}_\mathrm{DEP}=\sqrt{\frac{1-p}{3}}\hat{\sigma}_{z}.\non
\ee
for depolarizing channel.

Concretely, we choose the noisy channel as $\Lambda_{0}^{5}=\varepsilon_\mathrm{DEP}^{\otimes5}$ for five-qubit code, and after error correction, the fidelity of the effective channel $F'_{\Lambda_{0}^{5}}$ is
\begin{equation}
\label{f0def}
F'_{\Lambda_{0}^{5}}=\frac{1}{27}(5 + 20p - 70 p^2 + 40 p^3 + 160 p^4 - 128 p^5).
\end{equation}
This is the same as the result by Reimpell and Werner~\cite{Reimpell}. In the next, we choose quantum channels $\Lambda_{1}^{5}=\varepsilon_\mathrm{BF}\otimes\varepsilon_\mathrm{BPF}\otimes\varepsilon_\mathrm{PF}\otimes\varepsilon_\mathrm{AD}\otimes\varepsilon_\mathrm{GAD}$ as a contrast, and after error correction, the fidelity of the effective channel $F'_{\Lambda_{1}^{5}}$ becomes
\be
\label{fdef}
F'_{\Lambda_{1}^{5}}&=&3p-4p^\frac{3}{2}-3p^2+4p^\frac{5}{2}+5p^3\non\\
&&-8p^\frac{7}{2}+4p^4+8p^\frac{9}{2}-8p^5.
\ee

\begin{figure}[tbhp]
\centering
\includegraphics[width=0.4 \textwidth]{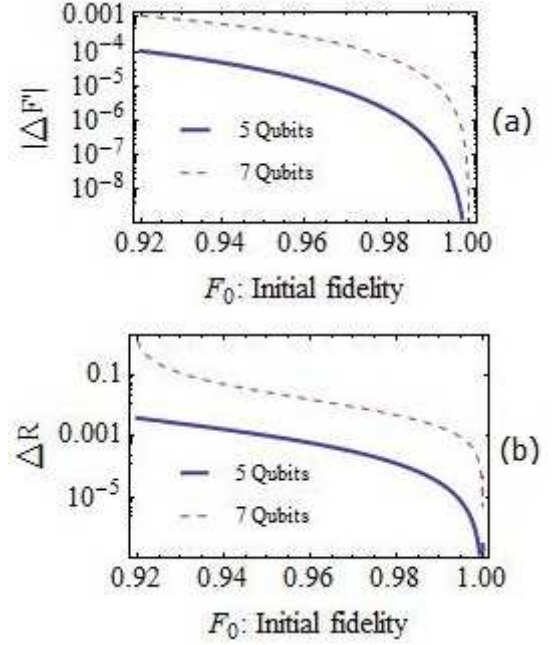}\\
\caption{(color online). (a) The fidelity gap of the effective channels between common errors and depolarizing errors after error correction for five-qubit (or seven-qubit) code. The numerical results for five-qubit code in Eq.~(\ref{f1def}) are shown with the solid line, while the ones for seven-qubit code in Eq.~(\ref{s1def}) are shown with the dashed line. (b) The relative deviations of fidelity gaps $\Delta R_{\Lambda_{1}^{5}}$ and $\Delta R_{\Lambda_{1}^{7}}$. The solid line shows the relative deviation of fidelity gap for five-qubit code in Eq.~(\ref{rddef}), while the dashed line shows the relative deviation of fidelity gap for seven-qubit code in Eq.~(\ref{rd7def}).}
\label{figure2}
\end{figure}

Then, define the fidelity gap $\Delta F'_{\Lambda_{1}^{5}}$ between specific common errors and depolarizing errors in QEC as
\begin{equation}
\label{fsdef}
\Delta F'_{\Lambda_{1}^{5}}=F'_{\Lambda_{1}^{5}}-F'_{\Lambda_{0}^{5}},
\end{equation}
and one can come to
\be
\label{f1def}
\Delta F'_{\Lambda_{1}^{5}}&=&\frac{1}{27}(-5+61p-108p^\frac{3}{2}-11p^2+108p^\frac{5}{2}\non\\
&&+95p^3-216p^\frac{7}{2}-52p^4+216p^\frac{9}{2}-88p^5).
\ee
If one chooses the initial fidelity $F_{0}=p=0.92$, it is shown in Fig.~\ref{figure2} that as $p$ is growing in the interval $[0.92,1]$, the fidelity gap $\Delta F'_{\Lambda_{1}^{5}}$ is always greater than $0$ and decreasing nearly to $0$ from $1.05533\times10^{-4}$.

In order to further compare the performances of five-qubit code between common error channels and depolarizing channels, the relative deviation of fidelity gap $\Delta R_{\Lambda_{1}^{5}}$ should firstly be defined
\begin{equation}
\label{rddef}
\Delta R_{\Lambda_{1}^{5}}=\frac{\Delta F'_{\Lambda_{1}^{5}}}{F'_{\Lambda_{0}^{5}}-F_{0}}=\frac{F'_{\Lambda_{1}^{5}}-F'_{\Lambda_{0}^{5}}}{F'_{\Lambda_{0}^{5}}-F_{0}},
\end{equation}
and this quantity  represents the relative deviation of fidelity gap from the improvement of fidelity in depolarizing channels for five-qubit code. The numerical results for Eq.~(\ref{rddef}) are shown in Fig.~\ref{figure2}.

\section{Exact performance of seven-qubit code for common error channels}
\label{v}

We will describe how seven-qubit code works against common errors in this section. We take the depolarizing channel as a contrast and then compare the performance of a common error channels with that of depolarizing channels. Similarly, the common errors chosen are bit flip error, bit-phase flip error, phase flip error, amplitude damping error, generalized amplitude damping error, and another two depolarizing channels for seven-qubit code.

For the initial channel fidelity $F_{0}=p$, take the noisy channel $\Lambda_{0}^{7}=\varepsilon_\mathrm{DEP}^{\otimes7}$ as a contrast for seven-qubit code, and after error correction, the fidelity of the effective channel $F'_{\Lambda_{0}^{7}}$ is
\be
\label{s0def}
F'_{\Lambda_{0}^{7}}&=&\frac{1}{729}(154+350p-1491p^2+2296p^3\non\\
&&+140p^4-4368p^5+8512p^6-4864p^7).
\ee

For the noisy evolution $\Lambda_{1}^{7}=\varepsilon_\mathrm{BF}\otimes\varepsilon_\mathrm{BPF}\otimes\varepsilon_\mathrm{PF}\otimes\varepsilon_\mathrm{AD}\otimes\varepsilon_\mathrm{GAD}\otimes\varepsilon_\mathrm{DEP}\otimes\varepsilon_\mathrm{DEP}$, the fidelity of the effective channel after error correction $F'_{\Lambda_{1}^{7}}$ becomes
\be
\label{sdef}
F'_{\Lambda_{1}^{7}}&=&\frac{1}{9}(3+5p^\frac{1}{2}-15p-4p^\frac{3}{2}+46p^2-28p^\frac{5}{2}\non\\
&&-136p^3+157p^\frac{7}{2}+347p^4-618p^\frac{9}{2}-48 p^5\non\\
&&+576p^\frac{11}{2}-244p^6-16p^\frac{13}{2}-16p^7).
\ee

Meanwhile, for the noisy evolution $\Lambda_{1}^{7}=\varepsilon_\mathrm{BF}\otimes\varepsilon_\mathrm{BPF}\otimes\varepsilon_\mathrm{PF}\otimes\varepsilon_\mathrm{AD}\otimes\varepsilon_\mathrm{GAD}\otimes\varepsilon_\mathrm{DEP}\otimes\varepsilon_\mathrm{DEP}$ we have chosen, similar to Eq. (\ref{fsdef}), the fidelity gap between the specific common errors and depolarizing errors after error correction is defined as
\be
\label{s1def}
\Delta F'_{\Lambda_{1}^{7}}&=&\frac{1}{729}(89+405p^{\frac{1}{2}}-1565p-324p^\frac{3}{2}+5217p^2\non\\
&&-2268p^\frac{5}{2}-13312p^3+12717p^\frac{7}{2}+27967p^4\non\\
&&-50058p^\frac{9}{2}+480p^5+46656p^\frac{11}{2}-28276p^6\non\\
&&-1296p^\frac{13}{2}+3568p^7).
\ee
If one chooses the initial fidelity $F_{0}=p=0.92$, it is shown in Fig.~\ref{figure2} that as $p$ is growing in the interval $[0.92,1]$, the fidelity gap $\Delta F'_{\Lambda_{1}^{7}}$ is smaller than $0$ and its absolute value $|\Delta F'_{\Lambda_{1}^{7}}|$ is decreasing nearly to $0$ from $1.11788\times10^{-3}$.

In order to further compare the performance of seven-qubit code between common error channels and depolarizing channels, the relative deviation of fidelity gap $\Delta R_{\Lambda_{1}^{7}}$ can also be defined
\begin{equation}
\label{rd7def}
\Delta R_{\Lambda_{1}^{7}}=\frac{|\Delta F'_{\Lambda_{1}^{7}}|}{F'_{\Lambda_{0}^{7}}-F_{0}},
\end{equation}
and this quantity represents the relative deviation of fidelity gap from improvement of fidelity in depolarizing channels for seven-qubit code. The numerical results for~Eq.~(\ref{rd7def}) are depicted in Fig.~\ref{figure2}.

From Fig.~\ref{figure2}, it is aware that for the noisy evolutions $\Lambda_{1}^{5}$ and $\Lambda_{1}^{7}$ above, the fidelity gap $|\Delta F'_{\Lambda_{1}^{7}}|$ for seven-qubit code is about $10$ times greater than $|\Delta F'_{\Lambda_{1}^{5}}|$ for five-qubit code. Meanwhile, relative deviation of fidelity gap $\Delta R_{\Lambda_{1}^{7}}$ for seven-qubit code is about $100$ times greater than $\Delta R_{\Lambda_{1}^{5}}$ for five-qubit code. Therefore, one can believe that five-qubit code is more effective than seven-qubit code in the case above. However, if the errors order change and become even arbitrary, does five-qubit code still work well? This is our following discussion.

\section{Five-qubit code agaist arbitrary error channels}
\label{vi}

In this section, the numerical calculation for five-qubit code is carried out where error channels are arbitrary, and this is different from Ref.~\cite{Huang}, where the error on each qubit is the same. Let \{$\bar{A}_m$\} be a set of Kraus operators, and introduce
an arbitrary $2\times2$ unitary transformation
\be
U_2(\theta,\phi)=\left(\begin{array}{cc}
\cos\frac{\theta}{2} & \sin\frac{\theta}{2}\exp[-i\phi]\\
-\sin\frac{\theta}{2}\exp[i\phi] & \cos\frac{\theta}{2}
\end{array}\right),\non
\ee
and another set of operators \{$A_m$\} can be introduced as
\begin{equation}
\label{mmdef}
A_m=U_2(\theta,\phi)\bar{A}_mU^{\dag}_2(\theta,\phi),
\end{equation}
where $\theta$ and $\phi$ are two free parameters. With three free parameters $\alpha$, $\beta$, and $\gamma$, the four operators $\bar{A}_m$ can be explicitly expressed as
\be
\bar{A}_1&=&\left(\begin{array}{cc}
\cos\alpha & 0\\
0 & \sin\beta\cos\gamma
\end{array}\right),\ 
\bar{A}_2=\left(\begin{array}{cc}
0 & 0\\
\sin\alpha\sin\gamma & 0
\end{array}\right),\non\\
\bar{A}_3&=&\left(\begin{array}{cc}
0 & \sin\beta\sin\gamma\\
0 & 0
\end{array}\right),
\bar{A}_4=\left(\begin{array}{cc}
\sin\alpha\cos\gamma & 0\\
0 & \cos\beta
\end{array}\right).\non
\ee
The detailed discussions about this model can be found in Ref.~\cite{Huang}.

As a constrain, fidelity for each qubit channel is chosen to be the same, and the free parameters $\theta$, $\phi$, $\alpha$, $\beta$ are arbitrary.  Then, each channel for the five qubits is generated independently from the arbitrary error model over $10^{5}$ times.  The $N$-th error process can be written as $\Lambda(N)=\varepsilon_{1}(N)\otimes\varepsilon_{2}(N)\otimes\varepsilon_{3}(N)\otimes\varepsilon_{4}(N)\otimes\varepsilon_{5}(N)$, and $\varepsilon_{1}(N),\varepsilon_{2}(N),\varepsilon_{3}(N),\varepsilon_{4}(N),\varepsilon_{5}(N)$ are generated independently based on Eq. (\ref{mmdef}). If five-qubit code works well against arbitrary errors, concatenated five-qubit code can be efficient either, and therefore, we just need to perform the first level error correction here. We choose the initial channel fidelity $F_{0}=$ {0.9, 0.91, 0.92, 0.93, 0.94, 0.945, 0.95, 0.96, 0.97, 0.98, 0.99, 0.992, 0.9993}.

\begin{figure}[tbph]
\centering
\includegraphics[width=0.4 \textwidth]{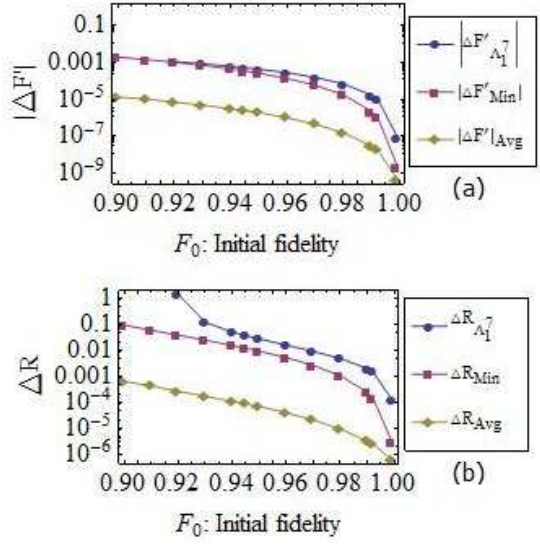}\\
\caption{(Color online). (a) The yellow line with dots shows the average fidelity gap of five-qubit code against arbitrary errors and depolarizing errors. The red line with dots shows the minmum fidelity gap of five-qubit code against arbitrary errors and depolarizing errors. The blue line with dots shows the fidelity gap of seven-qubit code against common chosen errors and depolarizing errors. (b) The yellow line with dots shows the relative deviation of the average fidelity gap when five-qubit code against arbitrary errors and depolarizing errors. The red line with dots shows the relative deviation of the min fidelity gap when five-qubit code against arbitrary errors and depolarizing errors. The blue line with dots shows the relative deviation of the fidelity gap when seven-qubit code against common chosen errors and depolarizing errors.}\label{figure3}
\end{figure}

After error correction with five-qubit code, based on Eq. (\ref{fsdef}), we employ the minmum fidelity of the effective channel $F'_\mathrm{min}$ to define the fidelity gap $\Delta F'_\mathrm{min}$, which shows the worst performance in error correction for five-qubit code,
\begin{equation}
\label{v1def}
\Delta F'_\mathrm{min}=F'_\mathrm{min}-F'_{\Lambda_{0}^{5}}.
\end{equation}
Since $\Delta F'_\mathrm{min}<0$, in Fig.~\ref{figure3}~(a) and Table~\ref{t1} we have taken its absolute value $|\Delta F'_\mathrm{min}|$ for convenience. Here $F'_{\Lambda_{0}^{5}}$ can be calculated by Eq. (\ref{f0def}).

Also based on Eq. (\ref{fsdef}), another fidelity gap $|\Delta F'|_\mathrm{avg}$ can be defined to show the average disparity of five-qubit code when applied in arbitrary error channels and depolarizing channels. Since the values of $N$-th fidelity gap $\Delta F'_N = F'_N - F'_{\Lambda^5_0}$ can be both positive and negative, we take the absolute value of the fidelity gap $|\Delta F'|_\mathrm{avg}$ as
\begin{equation}
\label{v2def}
|\Delta F'|_\mathrm{avg}=\frac{1}{N}\sum_{N}|F'_{N}-F'_{\Lambda_{0}^{5}}|.
\end{equation}
where $N$ is the time the calculation has be performed, and can reach over $10^{5}$, $F'_{N}$ means the $N$-th fidelity of the effective channel after error correction, and $F'_{\Lambda_{0}^{5}}$ can be obtained in Eq.~(\ref{f0def}) either. The numerical results for these quantities are shown in Fig.~\ref{figure3}~(a) and listed in Table~\ref{t1}.

Meanwhile, when initial fidelity $F_{0}=0.92$, $0.93$, $0.94$, $0.945$, $0.95$, $0.96$, $0.97$, $0.98$, $0.99$, $0.992$, $0.9993$, based on Eq.~(\ref{s1def}), the absolute values of fidelity gap for seven-qubit code $|\Delta F'_{\Lambda_{1}^{7}}|$ are listed in Table~\ref{t1}.
\begin{table}
\caption{The worst performance fidelity gap and the average fidelity gap of five-qubit code between an arbitrary error channel and a depolarizing channel, and the fidelity gap of seven-qubit code between common error channels and depolarizing channels. }\label{t1}
\centering\begin{tabular}{p{1.2cm} p{2.45cm} p{2.45cm} p{2.45cm} }
\hline\hline\noalign{\smallskip}
    $F_{0}$ & $|\Delta F'_\mathrm{min}|$ & $|\Delta F'|_\mathrm{avg}$ & $|\Delta F'_{\Lambda_{1}^{7}}|$\\
\hline\noalign{\smallskip}
   $0.9$  & $1.95185\times10^{-3}$     & $1.32228\times10^{-5}$  &None \\
   \hline\noalign{\smallskip}
   $0.91$  & $1.44212\times10^{-3}$    & $1.04286\times10^{-5}$   &None \\
   \hline\noalign{\smallskip}
   $0.92$  & $1.02643\times10^{-3}$    & $7.09649\times10^{-6}$  & $1.11788\times10^{-3}$  \\
    \hline\noalign{\smallskip}
   $0.93$  & $6.96773\times10^{-4}$   & $4.93955\times10^{-6}$   & $8.58032\times10^{-4}$  \\
    \hline\noalign{\smallskip}
   $0.94$  & $4.44576\times10^{-4}$    & $3.10616\times10^{-6}$  & $6.30686\times10^{-4}$   \\
    \hline\noalign{\smallskip}
   $0.945$  & $3.44677\times10^{-4}$    & $2.49656\times10^{-6}$  & $5.29647\times10^{-4}$   \\
    \hline\noalign{\smallskip}
   $0.95$  & $2.60648\times10^{-4}$    & $1.95867\times10^{-6}$   & $4.37225\times10^{-4}$  \\
    \hline\noalign{\smallskip}
   $0.96$  & $1.35187\times10^{-4}$    & $1.00864\times10^{-6}$  & $2.78688\times10^{-4}$   \\
    \hline\noalign{\smallskip}
   $0.97$  & $5.77680\times10^{-5}$    & $4.53211\times10^{-7}$   & $1.55730\times10^{-4}$  \\
    \hline\noalign{\smallskip}
   $0.98$  & $1.73357\times10^{-5}$    & $1.53001\times10^{-7}$   & $6.85675\times10^{-5}$  \\
    \hline\noalign{\smallskip}
   $0.99$  & $2.19452\times10^{-6}$    & $2.96438\times10^{-8}$   & $1.69308\times10^{-5}$  \\
    \hline\noalign{\smallskip}
   $0.992$  & $1.12642\times10^{-6}$    & $1.86676\times10^{-8}$  & $1.08047\times10^{-5}$   \\
   \hline\noalign{\smallskip}
   $0.9993$  & $1.94983\times10^{-9}$    & $4.21489\times10^{-10}$  & $8.17661\times10^{-8}$   \\
\hline\hline
\end{tabular}
\end{table}

\begin{table}
\caption{The relative deviation of the worst fidelity gap and the average fidelity gap in five-qubit code, and the relative deviation fidelity gap in seven-qubit code.}\label{t2}
\centering\begin{tabular}{p{1.2cm} p{2.45cm} p{2.45cm} p{2.45cm} }
\hline\hline\noalign{\smallskip}
    $F_{0}$ & $\Delta R_\mathrm{min}$ & $\Delta R_\mathrm{avg}$ & $\Delta R_{\Lambda_{1}^{7}}$\\
\hline\noalign{\smallskip}
   $0.9$  & $9.52501\times10^{-2}$     & $6.45269\times10^{-4}$  &None \\
   \hline\noalign{\smallskip}
   $0.91$  & $5.99347\times10^{-2}$    & $4.33414\times10^{-4}$   &None \\
   \hline\noalign{\smallskip}
   $0.92$  & $3.84932\times10^{-2}$    & $2.66134\times10^{-4}$  & $1.45506$  \\
    \hline\noalign{\smallskip}
   $0.93$  & $2.47053\times10^{-2}$   & $1.75141\times10^{-4}$   & $0.119470$  \\
    \hline\noalign{\smallskip}
   $0.94$  & $1.55591\times10^{-2}$    & $1.08709\times10^{-4}$  & $5.16485\times10^{-2}$   \\
    \hline\noalign{\smallskip}
   $0.945$  & $1.21850\times10^{-2}$    & $8.82583\times10^{-5}$  & $3.74590\times10^{-2}$   \\
    \hline\noalign{\smallskip}
   $0.95$  & $9.42051\times10^{-3}$    & $7.07914\times10^{-5}$   & $2.79574\times10^{-2}$  \\
    \hline\noalign{\smallskip}
   $0.96$  & $5.32710\times10^{-3}$    & $3.97459\times10^{-5}$  & $1.61721\times10^{-2}$   \\
    \hline\noalign{\smallskip}
   $0.97$  & $2.67621\times10^{-3}$    & $2.09959\times10^{-5}$   & $9.30149\times10^{-3}$  \\
    \hline\noalign{\smallskip}
   $0.98$  & $1.07176\times10^{-3}$    & $9.45914\times10^{-6}$   & $4.93263\times10^{-3}$  \\
    \hline\noalign{\smallskip}
   $0.99$  & $2.43240\times10^{-4}$    & $3.28571\times10^{-6}$   & $2.01035\times10^{-3}$  \\
    \hline\noalign{\smallskip}
   $0.992$  & $1.52812\times10^{-4}$    & $2.53247\times10^{-6}$  & $1.54728\times10^{-3}$   \\
   \hline\noalign{\smallskip}
   $0.9993$  & $2.80508\times10^{-6}$    & $6.06365\times10^{-7}$  & $1.18156\times10^{-4}$   \\
\hline\hline
\end{tabular}
\end{table}

In the following, to show the relative deviation of fidelity gap for five-qubit code, according to Eq. (\ref{rddef}) and Eq. (\ref{v1def}), one can define the relative deviation of the min fidelity gap $\Delta R_\mathrm{min}$,
\begin{equation}
\label{v11def}
\Delta R_\mathrm{min}=\frac{|\Delta F'_\mathrm{min}|}{F'_{\Lambda_{0}^{5}}-F_{0}}.
\end{equation}
The numerical results for $\Delta R_\mathrm{min}$ are shown in Fig.~\ref{figure3}~(b) and listed in Table~\ref{t2}.

Similarly, according to Eq.~(\ref{v2def}), the relative deviation of the average fidelity gap $\Delta R_\mathrm{avg}$ can be defined as
\begin{equation}
\label{v22def}
\Delta R_\mathrm{avg}=\frac{|\Delta F'|_\mathrm{avg}}{F'_{\Lambda_{0}^{5}}-F_{0}},
\end{equation}
and the results are also shown in Fig.~\ref{figure3}~(b) and listed in Table~\ref{t2}. Moreover, when initial fidelity $F_{0}=0.92$, $0.93$, $0.94$, $0.945$, $0.95$, $0.96$, $0.97$, $0.98$, $0.99$, $0.992$, $0.9993$, via Eq. (\ref{rd7def}), the relative deviation for seven-qubit code $\Delta R_{\Lambda_{1}^{7}}$ is listed in Table~\ref{t2}.

As shown in Fig.~\ref{figure3}~(a) and listed in Table \ref{t1}, when the initial fidelity $F_{0}=0.9$, the worst performance fidelity gap between arbitrary errors and depolarizing errors $|\Delta F'_\mathrm{min}|=1.95185\times10^{-3}$, the average performance fidelity gap $|\Delta F'|_\mathrm{avg}=1.32228\times10^{-5}$, and they are both decreasing to nearly $0$ with almost the same speed as the initial fidelity increases, and in the process, $|\Delta F'_\mathrm{min}|$ keeps nearly $100$ times greater than $|\Delta F'|_\mathrm{avg}$.  As shown in Fig.~\ref{figure3}~(b) and listed in Table \ref{t2}, when the initial fidelity $F_{0}=0.9$, the relative deviation of the min fidelity gap $\Delta R_\mathrm{min}=9.52501\times10^{-2}$, the relative deviation of the average fidelity gap $\Delta R_\mathrm{avg}=6.45269\times10^{-4}$, and they are both decreasing to nearly $0$ with almost the same speed as the initial fidelity increases.

Meanwhile, when comparing the worst performance of five-qubit code with the performance of seven-qubit code in common error channels $\Lambda_{1}^{7}=\varepsilon_\mathrm{BF}\otimes\varepsilon_\mathrm{BPF}\otimes\varepsilon_\mathrm{PF}\otimes\varepsilon_\mathrm{AD}
\otimes\varepsilon_\mathrm{GAD}\otimes\varepsilon_\mathrm{DEP}\otimes\varepsilon_\mathrm{DEP}$, it is found that even the worst performance of five-qubit code is better than that (not the worst one) of seven-qubit code for both $|\Delta F'|$ and $\Delta R$. Furthermore, the worst performance fidelity gap is the fidelity gap between bit-flip error channels and depolarizing error channels except an anomalous data in Table \ref{t1} where initial fidelity is $0.9993$. [With the same initial fidelity $0.9993$, the fidelity gap between a bit-flip error and a depolarizing errors is $7.61556\times10^{-10}$]. However, this does not matter if the fidelity of effective channel after error correction in bit-flip error channels is taken as the threshold for five-qubit code, where the initial channel fidelity is required to be below $0.992$.

\section{Remarks and discussion}
\label{vii}

In section~\ref{iv} and section~\ref{v}, it is shown that five-qubit code is more efficient than seven-qubit code in a complicated noise environment via an analytic example in this work, and in section~\ref{vi}, the numerical calculation for five-qubit code has been carried out when error models are arbitrary. We find the worst performance fidelity gap between arbitrary error channels and depolarizing channels is sufficient small, and it is a little better than the performance of seven-qubit code in error channels. Meanwhile, the worst performance fidelity gap is the fidelity gap between bit-flip error channels and depolarizing error channels in most cases, say when initial channel fidelity below $0.992$.

Based on the analysis in the present work, the following results can be summarized: (1) In the preparation of QEC, only the channel fidelity is required in the construction of the QECCs. (2) When five-qubit code is employed in bit-flip error channels, where the initial channel fidelity is below $0.992$, one can take the fidelity of effective channel as the threshold in the construction of QEC. (3) When designing QEC with concatenated five-qubit code, it is not necessary to adjust the protocol even when errors in each qubit are time-varying or level-varying.

In recent works, suppression of coherent error in multiqubit entangling gates in trapped ion systems has been introduced~\cite{D. Hayes}, and  the effect of coherent errors on the logical error rate of the Steane $[[7,1,3]]$ QECC has also been studied~\cite{M. Guti¨¦rrez}. Quite recently, failure distributions of coherent and stochastic error models in QEC have been compared~\cite{Barnes}. Therefore, the application of five-qubit code in suppression of coherent error will be our future consideration.  When protecting a quantum system from coherent error, the effective channel of every qubit in the system may not be the same in general case, and based on the results obtained in this work, the five-qubit code may be a good choice for the preservation of the system.

\section*{acknowledgements}
This work was supported by the National Natural Science Foundation of China under Grant No.~11405136, and the Fundamental Research Funds for the Central Universities under Grant No.~2682016CX059.

\end{document}